\documentclass[12pt]{iopart}
\usepackage{iopams}
\newtheorem{Theorem}{Theorem}
\def\cF{{\cal F}} \def\cH{{\cal H}}   \def\cZ{{\cal Z}}
\def\bG{\bi G} \def\bK{\bi K} \def\bZ{\bi Z}
\def\bn{{\bi n}}    \def\bs{{\bi s}}
\def\bone{{\bf 1}}
\def\ga{\alpha} \def\gg{\gamma} \def\gd{\delta}
\def\gee{\epsilon} \def\gl{\lambda} \def\gs{\sigma}
\def\gk{\kappa}
\def\gD{\Delta} \def\gG{\Gamma}
\def\mutm{{\mu t_\mu m}} \def\mut{{\mu t_\mu }}
\def\ket#1{\mid~\!\!\!{#1}~\!\!\rangle}
\def\bra#1{\langle~\!\!{#1}~\!\!\!\mid}
\def\braket#1#2{\langle~\!{#1}~\!\mid~\!{#2}~\!\rangle}
\def\mod#1{\mid #1\mid}
\def\d={\buildrel \rm def \over =}
\begin{document}\jl{1}
\title[ON THE DIRAC-HEISENBERG MODEL]{GROUP PROJECTOR GENERALIZATION OF DIRAC-HEISENBERG MODEL}
\author{M. Damnjanovi\'c\footnote{e-mail: yqoq@afrodita.rcub.bg.ac.yu}}
\address{Faculty of Physics, P.O.Box
368, 11001 Beograd, Serbia, Yugoslavia,
http://www.ff.bg.ac.yu/qmf/qsg\_e.htm}

\date{\today}
\begin{abstract}
The general form of the operators commuting with the ground
representation (appearing in many physical problems within single
particle approximation) of the group is found. With help of the
modified group projector technique, this result is applied to the
system of identical particles with spin independent interaction,
to derive the Dirac-Heisenberg hamiltonian and its effective space
for arbitrary orbital occupation numbers and arbitrary spin. This
gives transparent insight into the physical contents of this
hamiltonian, showing that formal generalizations with spin greater
than 1/2 involve nontrivial additional physical assumptions.
\end{abstract}
\submitted 

\baselineskip=0.3in
\section{Introduction}\label{INTRO}
Considering systems of identical electrons interacting by Coulomb
forces only, Dirac found \cite{DIRAC} that the effective
hamiltonian can be expressed in the spin space only:
$H=U+\sum_{k<l}J_{kl}\bs_k\bdot\bs_l$, where $\bs_i$ is vector of
the Pauli matrices related to the spin in the $i$-th site. The aim
of this paper is to present rigorous derivation of the
Dirac-Heisenberg hamiltonian for any spin, within the framework of
the original physical assumptions. This means that arbitrary spin
independent interaction of the identical particles is considered.
Then, due to the perturbative approach, the hamiltonian is
approximately reduced in the subspaces of the orbital state space
spanned by the vectors with the same occupation number. Such
subspace carries special induced type representation (ground
representation) of the permutational group, commuting with the
reduced hamiltonian. The general form of the operator commuting
with the ground representation is derived in section \ref{Sinv},
and the result is applied to the considered hamiltonian in section
\ref{Sdhind}, yielding its form in the orbital many particle
factor space. Finally, the wanted form of the hamiltonian is
obtained in the section \ref{SHasym}, by the restriction to the
relevant (symmetrized or antisymmetrized) subspace of the total
space. This step is based on the modified group projector
technique for the induced representations.

The result generalizes the original derivation with respect to
spin and occupation numbers. Nevertheless, the physical framework
remains the same, in contrast to the formal generalizations
appearing in various theories of magnetic materials
\cite{MANOUSAKIS}, when in the Dirac-Heisenberg hamiltonian only
the values of spin and the interaction coefficients are
appropriately modeled.

The rest of the introduction gives the necessary reminder on the
modified group projector technique. Let $D({\bG})$ be
representation of the group ${\bG}$ in the space $\cH_{D}$,
decomposing into the irreducible components $D^{(\mu )}({\bG})$ as
$D({\bG})=\oplus_{\mu =1}^ra_\mu D^{(\mu )}({\bG})$ ($a_\mu$ is
the frequency number of $D^{(\mu )}({\bG})$). The symmetry adapted
\cite{ELDOB,JABOON} (or standard) basis $\{\ket{\mutm}|\mu
=1,\dots , r;\ t_\mu =1,\dots ,a_\mu ;\ m=1,\dots ,|\mu|\}$ ($|\mu
|$ denotes the dimension of $D^{(\mu )}({\bG})$) in $\cH_{D}$ is
defined by the following condition:
\begin{equation}\label{SAB}
D(g)\ket{\mutm}=\sum_{m'=1}^{|\mu |}D^{(\mu
)}_{m'm}(g)\ket{\mutm'}.
\end{equation}
To find this basis \cite{YI-GP}, the auxiliary representation
$\gG^\mu({\bG})\d=D({\bG})\otimes D^{(\mu)*}({\bG})$ in the space
$\cH_D\otimes\cH^{(\mu)^*}$ is constructed for each irreducible
component $D^{(\mu)}({\bG})$ (with $a_\mu>0$) of $D({\bG})$. Here,
$D^{(\mu)*}({\bG})$ is the dual representation of
$D^{(\mu)}({\bG})$; in fact, it is the conjugated one, since the
finite permutational groups and their unitary representations are
considered. The range of the modified projector
$G(\gG^\mu)\d=\frac{1}{\mod{{\bG}}}\sum_g\gG^\mu({\bG})$ is the
($a_\mu$ dimensional) subspace $\cF^\mu$ of the fixed points for
the representation $\gG^\mu({\bG})$. For the arbitrary basis
$\{\ket{\mut}~|~\mu=1,\dots,a_\mu\}$ of $\cF^\mu$, the subbasis
$\{\ket{\mutm}~|~m=1,\dots,\mod{\mu}\}$ is found by the partial
scalar product with the standard vectors $\ket{\mu m}$ of the
irreducible representation:
\begin{equation}\label{SABM}
\ket{\mutm}=\braket{\mu m}{\mut}.
\end{equation}
If ${\bG}$ is the symmetry group of the hamiltonian $H$ (thus
$[D(g),H]=0$ for each $g\in {\bG}$), then taking  for $\ket{\mu
t_\mu}$ an eigen basis of $H\otimes I_\mu$ ($I_\mu$ is the
identity in $\cH^{(\mu)^*}$), eq. \eref{SABM} gives the symmetry
adapted eigen basis for $H$: $H\ket{\mutm}=E_{\mu
t_\mu}\ket{\mutm}$.

The representations involved in the paper are of the induced type.
Precisely, let ${\bK}$ be subgroup of ${\bG}$ with the transversal
${\bZ}=\{z_t~|~t=0,\dots,\mod{{\bZ}}-1\}$ ($z_0$ is the identity
of the group, $\mod{{\bZ}}=\frac{\mod{{\bG}}}{\mod{{\bK}}}$). Then
$D({\bG})=\gD(\bG)\otimes d({\bG})$, where
$\gD(\bG)=\gD'({\bK}\uparrow {\bG})$ is induced representation and
$d({\bG})$ is some other representation of ${\bG}$. In this case
the modified projectors can be reduced \cite{YI-IND} to the
subgroup modified projector $K(\gg^\mu)$ for the representation
$\gg^\mu({\bK})=\gD'({\bK})\otimes d({\bG}\downarrow {\bK})\otimes
D^{(\mu)*}({\bG}\downarrow {\bK})$ in
$\cH_{\gg^\mu}=\cH_{\gD'}\otimes\cH_d\otimes\cH^{(\mu)^*}$:
\begin{equation}\label{BASIC}
G(\gG^\mu)=B^\mu\{E^{00}\otimes K(\gg^\mu)\} B^{\mu^\dagger}.
\end{equation}
Here, $B^\mu\d=\frac{1}{\sqrt{|{\bZ}|}}\sum_{z_t}E^{t0}\otimes
I_{\gD'}\otimes d(z_t)\otimes D^{(\mu)^*}(z_t)$ is partial
isometry, and $E^{t0}=\ket{z_t}\bra{z_0}$ are
$|{\bZ}|$-dimensional square matrices with only one non-vanishing
element $(E^{t0})_{t0}=1$. It appears that the range of
$K(\gg^\mu)$ (the subspace in $\cH_{\gg^\mu}$) is the effective
space, while the effective hamiltonian is
$H^\mu=B^{\mu^\dagger}(H\otimes I_\mu)B^\mu K(\gg^\mu)$. Indeed,
the symmetry adapted eigen subbasis $\ket{\mutm}$ corresponding to
the irreducible representation $D^{(\mu)}({\bG})$ is found by
(\ref{SABM}) with the vectors
$\ket{\mut}=B^{\mu^\dagger}\ket{\mut}^0$, where $\ket{\mut}^0$ are
the eigen vectors of $H^\mu$ from the range of $K(\gg^\mu)$:
$H^\mu\ket{\mut}^0=E_{\mut}\ket{\mut}^0$.

\section{Invariants of ground representations}\label{Sinv}

If ${\bK}$ is subgroup in the finite group ${\bG}$, its left
transversal ${\bZ}$ gives the coset partition
${\bG}=\sum_tz_t{\bK}$. Therefore, to each element $g\in{\bG}$
corresponds one element $\overline{g}$ of ${\bZ}$: there are
uniquely defined $k\in{\bK}$ and $z_t\in {\bZ}$, such that
$g=z_tk$, and $z_t$ is denoted by $\overline{g}$. Together with
this coset decomposition, the subgroup ${\bK}$ gives the
double-coset decomposition \cite{ALTMAN,McK,STERN} of ${\bG}$ over
${\bK}$: ${\bG}=\sum_\gl {\bK}z_\gl {\bK}$. Each double-coset
decomposes onto one or more cosets, ${\bK}z_\gl {\bK}=\sum_mz_{\gl
m}{\bK}$. Thus, the double-coset representatives can be chosen
among the elements of the transversal ${\bZ}$. The double-coset
decomposition enables to define for each $g\in {\bG}$ its
double-coset representative $z_\gl$ by $g=kz_\gl k'$ ($k,k'\in
{\bK}$), denoted also as $\overline{\overline{g}}$.

Furthermore, each $g\in {\bG}$, and $z_m\in {\bZ}$ define uniquely
$z_s\in {\bZ}$ and $k\in {\bK}$, such that $gz_m=z_sk$. Obviously,
with the above notational convention, $z_s=\overline{gz_m}$, and
the left (permutational) action of ${\bG}$ over ${\bZ}$ becomes
$g:z_m\mapsto\overline{gz_m}$. This action is faithfully
represented by the linear operators of the left ground
representation $L({\bG})=\bone({\bK}\uparrow {\bG})$ in the
$\mod{{\bZ}}$-dimensional vector space, $\cZ$: each element of
$z_m\in {\bZ}$ is mapped to the basis vector $\ket{z_m}$. The
operators of ${\bG}$ are defined by
$L(g)\ket{z_m}\d=\ket{\overline{gz_m}}$, i.e.
$L(g)=\sum_m\ket{\overline{gz_m}}\bra{z_m}$.  The homomorphism
condition $L(gg')=L(g)L(g')$ is easily checked.

Also, the right multiplication $z_mg=z_sk$ introduces the
"right" operators $R(g)$: $\bra{z_m}R(g)\d=\bra{\overline{z_mg}}$, or
$R(g)=\sum_m\ket{z_m}\bra{\overline{z_mg}}$. These operators form
antirepresentation ($R(g)R(g')=R(g'g)$) if and only if ${\bK}$ is
invariant subgroup. Since $\overline{z_sz_mk}=\overline{z_sz_m}$, it
turns out that $R(z_mk)=R(z_m)$ for each $k\in {\bK}$, i.e. that the
mapping $g\mapsto R(g)$ is function over the cosets of ${\bK}$.

All the operators $L(g)$ and $R(g)$ are in the basis
$\{\ket{z_m}\}$ given by the real matrices, with elements 0 or 1.
Especially, for the unitary (in fact orthogonal) matrices $L(g)$
this yields $L(g)^T=L(g^{-1})$.

Now, the condition that the operator $A$ acting in $\cZ$ is
invariant of ${\bG}$ means that $[A,L(g)]=0$ for each $g$ in
${\bG}$. Such an operator has very special form.
\begin{Theorem}\label{Tinv}
Any invariant operator $A$ in $\cZ$ is of the form $A=\sum_{g\in
{\bG}}\ga(\overline{\overline{g}})R(g)$, where
$\ga(\overline{\overline{g}})$ is function over double-cosets of
${\bK}$ in ${\bG}$ (i.e. these constants can be independently
chosen one for each double-coset).
\end{Theorem}
The proof consists of two parts. At first, the commutation with
the operators $L(z_m)$ representing the transversal is used:
because of $L(z_m)\ket{z_0}=\ket{\overline{z_mz_0}}=\ket{z_m}$,
one has
$A=\sum_{m,n}\bra{m}A\ket{n}\ket{m}\bra{n}=
\sum_{mn}\bra{z_0}L^T(z_m)A\ket{n}\ket{m}\bra{n}.$
Since $z_m^{-1}$ is also an element of ${\bG}$, it commutes with $A$,
and
$$A=\sum_{mns}\bra{z_0}A\ket{z_s}\bra{z_s}L^T(z_m)\ket{z_n}\ket{z_m}\bra{z_n}=
\sum_{ms}A^{0s}\ket{z_m}\bra{\overline{z_mz_s}},$$
giving finally $A=\sum_sA^{1s}R(z_s)$. Consequently, the matrix of
the invariant $A$ is completely determined by its first row.
Secondly, the subgroup elements are employed; for each double-coset
representative $z_\gl$ and each element $k\in {\bK}$ it holds
$$A^{0\gl}=\bra{z_0}A\ket{z_\gl}=\bra{z_0}L(k)A\ket{z_\gl}=
\bra{z_0}AL(k)\ket{z_\gl}=\bra{z_0}A\ket{\overline{kz_\gl}}.$$
When $k$ goes over ${\bK}$, all the elements $\overline{kz_\gl}$
go over the coset representatives of the whole double-coset of
$z_\gl$, meaning that the matrix elements $A^{0s}$ and $A^{0t}$
must be same if $z_t$ and $z_s$ are from the same double-coset.
Together with the previous conclusion this gives $A=\sum_\gl
A^{0\gl}\sum_mR(z_{\gl m})$. To complete the proof, it remains to
recall that the right operators are same for the elements of the
same coset.

From the theorem \ref{Tinv} immediately follows that the number of
linearly independent invariants is equal to the number of
double-cosets of ${\bK}$. Precisely, to each double-coset
represented by $z_\gl$, there corresponds the invariant
$A_\gl=\sum_{g\in {\bK}z_\gl {\bK}}R(g)$. In the special case,
when ${\bK}=\{e\}$, the trivial subgroup containing the identity
only, the ground representation is the regular representation of
the group; since in this case each element of the group is itself
one coset and one double-coset, there are exactly $\mod{{\bG}}$
independent invariants, each of them being one of the operators
$R(g)$ (in this case $R(g^{-1})$ form the right regular
representation of ${\bG}$, being equivalent to the left one
$L({\bG})$), and all the left operators commute with all the right
ones.

\section{Generalized Dirac-Heisenberg hamiltonian}\label{Sdhind}

Let $\cH=\cH_o\otimes\cH_s$ be the quantum mechanical state space
of some particle, where $\cH_o$ and $\cH_s$ are the orbital and
the spin factor spaces. Then, for the system of $N$ particles the
tensor powers $\cH_o^N\d=\cH_o\otimes\cdots\otimes\cH_o$ ($N$
times) and $\cH_s^N$ are constructed, and in the space
$\cH^N=\cH_o^N\otimes\cH_s^N$ the symmetric (bosons) or
antisymmetric (fermions) part are considered as the state space of
the total system. If $\{\ket{i}~|~i=1,\dots,\mod{o}\}$ is a basis
in $\cH_o$, then
$\{\ket{i_1,\dots,i_n}\d=\ket{i_1}\cdots\ket{i_N}~|~i_1,\dots,i_N=1,\dots,\mod{o}\}$
is a basis in $\cH_o^N$.  Each of this vectors defines the
occupation number vector $\bn=(n_1,\dots,n_{\mod{o}})$, with the
component $n_i$ showing the number of particles being in the state
$\ket{i}$.

Each permutation $\pi$ of the symmetric group $S_N$, is
represented by the operator $\gD(\pi)$, defined by the action
$\gD(\pi)\ket{i_1,\dots,i_N}\d=\ket{i_{\pi^{-1}1},\dots,i_{\pi^{-1}N}}$.
This action does not change the occupation number of the basis
vectors, and exactly the orbit of the action gives the set of the
basis vectors with the same occupation numbers. Thus, each orbit
is uniquely defined by the occupation number and spans the
subspace $\cH_{\bn}^N$  invariant for the representation
$\gD(S_N)$. Consequently, $\gD(S_N)$ is in $\cH^N_{\bn}$ reduced
to the representation $\gD_{\bn}(S_N)$. Its dimension is equal to
the order of the orbit with the occupation number $\bn$,
$\mod{\gD_{\bn}}=\frac{N!}{n_1!\cdots n_{\mod{o}}!}$, since the
stabilizer of the vector with the occupation number $\bn$ is
$S^{\bn}_N=S_{n_1}\otimes\cdots\otimes S_{n_{\mod{o}}}$, this is.
Note that, being induced from the trivial representation of the
stabilizer, $\gD_{\bn}(S_N)=\bone(S^{\bn}_N\uparrow S_N)$,
$\gD_{\bn}(S_N)$ is a ground representation \cite{ALTMAN}.

To summarize, the space $\cH_o^N$ is decomposed to the orthogonal
sum $\cH^N_o=\oplus_{\bn}\cH^N_{\bn}$. In each of these subspaces
acts the ground representation $\gD_{\bn}(S_N)$, and the partial
reduction of the representation $\gD(S_N)$ is obtained:
$\gD(S_N)=\oplus_{\bn}\gD_{\bn}(S_N)$.

Let $H$ be spin-independent hamiltonian of the system of $N$
identical particles. It is written in the form $H=H_1+H_2$, where
$H_1=\sum_{s=1}^N h_s$ is the noninteracting part. Here, $h_i$ is
one particle hamiltonian, i.e. the operator in the space $\cH_o$,
while $H_2=\sum_{s<t}V_{st}$ describes two-particle interaction.
Since $H$ commutes with the operators $\gD(S_N)$, all $h_s$ must
be equivalent: the full form of $h_s$ is the tensor product of the
identity operators in all the spaces except in the $s$-th one,
where the corresponding factor is same, e.g. $h$. Analogously, all
the operators $V_{st}$ are same except that their nontrivial
action is reduced to the different pair of spaces.

If the basis $\{\ket{i}\}$ is chosen as the eigen basis of $h$
(with the eigenvalues $\gee_i$), then the vectors of the subspace
$\cH^N_{\bn}$ are the eigenvectors of $H_1$ for the eigenvalue
$E_{\bn}=\sum_{i=1}^dn_i\gee_i$. Although this subspace need not
be invariant for $H_2$, the approximation $H_2\approx\oplus_\bn
H_{2\bn}$, with $H_{2\bn}=P_{\bn}H_2P_{\bn}$ ($P_{\bn}$ stands for
the projector onto $\cH^N_{\bn}$) enables the perturbative
approach, involving the eigen problems of the operators
$H_{2\bn}$. Since $H_2$ is invariant of $S_N$ in the whole space
$\cH^N_o$, the operators $H_{2\bn}$ are also $S_N$-invariants,
i.e. they commute with the corresponding representation
$\gD_{\bn}(S_N)$. Recalling that this is ground representation,
the theorem \ref{Tinv} gives the most general form
\begin{equation}\label{EH2gen}
H_{2\bn}=\sum_{\pi\in
S_N}\ga(\overline{\overline{\pi}})R_{\bn}(\pi).
\end{equation}
Here, $R_{\bn}(\pi)$ are the right operators of $\gD_{\bn}(\pi)$,
while the coefficients $\ga$ are equal for all the permutations
from the same double-coset of $S^{\bn}_N$.

Until now, only the orbital space $\cH^N_o$ has been considered,
since the hamiltonian acts trivially in the spin factors.
Nevertheless, the particles are identical, and either the
symmetrized or the antisymmetrized part of the total space $\cH^N$
is to be considered. The orbital occupation number decomposition
yields the decomposition of the total space:
$\cH^N=\oplus_{\bn}\cH^N_{\bn}\otimes\cH^N_s$.  Using arbitrary
basis $\{\ket{\gs}~|~\gs=1,\dots,2s+1\}$ in the single particle
spin space $\cH_s$, the representation $d(S_N)$ in $\cH^N_s$ is
defined analogously to $\gD(S_N)$ in $\cH^N_o$:
$d(\pi)\ket{\gs_1,\dots,\gs_N}\d=\ket{\gs_{\pi^{-1}1},\dots,\gs_{\pi^{-1}N}}$,
and in the total space the permutation $\pi$ is represented by the
operator $\gD(\pi)\otimes d(\pi)$. Obviously the subspaces
$\cH^N_{\bn}\otimes\cH^N_s$ are invariant for the action of
$\gD\otimes d$, and the modified group projector of the
irreducible representation $D^{(\mu)}(S_N)$
\begin{equation}\label{EMGPmu}\fl
S_N(\gD\otimes d\otimes D^{(\mu)*})= \frac{1}{N!}\sum_\pi
\gD(\pi)\otimes d(\pi) \otimes D^{(\mu)*}(\pi)=\oplus_\bn
S_N(\gD_\bn\otimes d\otimes D^{(\mu)*})
\end{equation}
independently treats each of these subspaces. Therefore, in each
subspace $\cH^N_{\bn}\otimes\cH^N_s$ there is the subspace
$\cH^{\mu}_{\bn s}$ corresponding to the representation
$D^{(\mu)}(S_N)$. It is spanned by the standard subbasis
$\{\ket{\mutm}\}$, obtained by \eref{SABM} from any basis
$\{\ket{\bn;\mut}\}$ of the range $\cF^\mu_\bn$ of the projector
$S_N(\gD_\bn\otimes d\otimes D^{(\mu)*})$; obviously,
$\cF^\mu_\bn$ is the intersection of
$\cH^N_{\bn}\otimes\cH^N_s\otimes\cH^{(\mu)^*}$ and the range
$\cF^\mu$ of the projector \eref{EMGPmu}. Especially, taking the
identity and the alternating representations,
$D^{(\pm)}(\pi)=(\pm)^\pi$ (as usual, $\pi$ in the exponent
denotes the parity of $\pi$) for $D^{(\mu)}(S_N)$, the projector
\eref{EMGPmu} becomes the symmetrizer and antisymmetrizer,
respectively; in these cases of one dimensional irreducible
representations $\cF^\pm_\bn$ is itself the subspace
$\cH^{\pm}_{\bn s}$.

\section{Restriction to the relevant subspace}\label{SHasym}

Since the involved representations are of the induced type, the
modified group projector technique isomorphically relates by eq.
\eref{BASIC} the subspace $\cF^\mu_\bn$ of
$\cH^N_{\bn}\otimes\cH^N_s\otimes\cH^{(\mu)^*}$ to the effective
subspace, being the range of the subgroup projector $\bK(\gg^\mu)$
in  $\cH^N_s\otimes\cH^{(\mu)^*}$ (since $\gD'(\bK)$ is trivial
representation). Of course, in the space
$\cH^N_s\otimes\cH^{(\mu)^*}$ acts the effective hamiltonian
$H^\mu_{2\bn}$, with the range contained in $\cF^\mu_\bn$.
Especially, for the physically important representations
$D^{(\pm)}(S_N)$, the effective hamiltonian $H^\pm_{2\bn}$ acts in
the spin space $\cH^N_s$, as well as $\bK(\gg^\pm)$.

Precisely, in the considered context ${\bG}=S_N$,
${\bK}=S_N^{\bn}$, $\gD'({\bK})=\bone(S_N^{\bn})$ and $d(S_N)$ is
the permutational representation in the spin space. Thus,
$$B^\mu_\bn=\frac{1}{\sqrt{\mod{{\bZ}}}}\sum_t\ket{z_t}\bra{z_0}\otimes
d(z_t)\otimes D^{(\mu)*}(z_t)$$
(omitted number 1 standing for $\gD'$). Then, skipping the factor
$E^{00}=\ket{z_0}\bra{z_0}$ (this only precisely gives the space
of action of $H^\mu_{2\bn}$), one finds:
$$B^{\mu^\dagger}_\bn(H_{2\bn}\otimes I_\mu)B^\mu_\bn
=\frac{1}{N!}\sum_\pi\sum_{p,t}\ga(\overline{\overline{\pi}})
\bra{z_t}R(\pi)\ket{z_p} d(z^{-1}_tz_p)\otimes
D^{(\mu)*}(z^{-1}_tz_p).$$
The matrix element of $R(\pi)$ is obviously
$\braket{\overline{z_t\pi}}{z_p}=\gd_{z_p,\overline{z_t\pi}}$
(Kronecker delta). When the sum over $\pi=z_q\gk$ is decomposed
onto the sums over transversal ($q$) and stabilizer ($\gk$), the
equality $\overline{z_tz_q\gk}=\overline{z_tz_q}$ for $\gk\in
S_N^{\bn}$ shows that all the terms are independent of $\gk$.
Thus:
$$B^{\mu^\dagger}_\bn(H_{2\bn}\otimes
I_\mu)B^\mu_\bn=\frac{1}{\mod{\bZ}}\sum_{q,t}\ga(\overline{\overline{z_q}})
d(z^{-1}_t\overline{z_tz_q})\otimes
D^{(\mu)*}(z^{-1}_t\overline{z_tz_q}).$$
Since the element $z^{-1}_t\overline{z_tz_q}$ can be written in
the form $z_q\gk'$ (i.e. it is from the coset represented by
$z_q$), multiplication by $S^\bn_N(d\otimes D^{(\mu)^*})$ gives:
$$H^\mu_{2\bn}=\frac{\mod{{\bZ}}}{N!}\sum_q\sum_{\gk}\ga(\overline{\overline{z_q}})
d(z_q\gk)\otimes D^{(\mu)*}(z_q\gk),$$
and finally,
\begin{equation}\label{EHmu}
H^\mu_{2\bn}=\frac{1}{n_1!\cdots
n_{\mod{o}}!}\sum_\pi\ga(\overline{\overline{\pi}}) d(\pi)\otimes
D^{(\mu)*}(\pi).
\end{equation}
This relation is in fact the most general form of the $\mu$-th
component of the permutational invariant hamiltonian acting in
$\cH^N_{\bn}\otimes \cH^N_s$, being trivial in $\cH^N_s$. Note
that this operator acts in the space isomorphic to the direct
product of $\cH^N_s$ and the space of the representation
$D^{(\mu)}(S_N)$. Still, the range of the projector
$K(\gg^\mu)=\frac{1}{n_1!\cdots
n_{\mod{o}}!}\sum_{\gk}\gg^\mu(\gk)$ is the effective part of this
space (its orthocomplement is from the kernel of $H^\mu_{2\bn}$).
Finally, let it be stressed again that the coefficients $\ga$ can
be deliberately chosen only one for each double-coset of
$S^{\bn}_N$.

Physically relevant are two simplifications. At first, as it has
been mentioned already, the irreducible representation
$D^{(\mu)}(S_N)$ is actually either the symmetric or the
antisymmetric one, giving:
\begin{equation}\label{EHpm}
H^{\pm}_{2\bn}=\frac{1}{n_1!\cdots n_{\mod{o}}!}\sum_\pi(\pm)^\pi
\ga(\overline{\overline{\pi}})d(\pi).
\end{equation}
In these cases, the effective space of $H^{\pm}_{2\bn}$ is
subspace in the spin space $\cH^N_s$.

The second one is that only two particle interaction are
considered, meaning that only the permutations of at most two
particles are involved in (\ref{EH2gen}). With $\tau_{kl}$
denoting the transposition of the particles $k$ and $l$,
expressions (\ref{EH2gen}) and (\ref{EHpm}) become
$H_{2\bn}=\ga(e)
I+\sum_{l<k}\ga(\overline{\overline{\tau_{lk}}})R_{\bn}(\tau_{lk})$,
and
\begin{equation}\label{EHpm2}
H^{\pm}_{2\bn}=\frac{1}{n_1!\cdots n_{\mod{o}}!}\left[\ga(e)
I\pm\sum_{k<l}
\ga(\overline{\overline{\tau_{kl}}})d(\tau_{kl})\right].
\end{equation}

\section{Concluding remarks}

Originally, the hamiltonian (\ref{EHpm2}) is derived \cite{DIRAC}
for the case when the orbital occupation numbers $n_i$ are at most
1, meaning that $S^{\bn}_N$ is the trivial subgroup $\{e\}$, and
therefore $\gD_{\bn}(S_N)$ is the regular ($N!$-dimensional)
representation of $S_N$. In this case the coefficients $\ga$ can
be chosen arbitrary, since each element of $S_N$ is itself one
double-coset. Further, in this case the group projector
$K(\gg^{\pm})$ is the identity operator in the space $\cH^N_s$,
meaning that the whole spin space is efficient.

Of course, for spin $s=1/2$ the transposition $\tau_{ij}$ is in
the space $H^N_s$ represented by the operator $d(\tau_{ij})=
\frac12(I+\bs_i\bdot\bs_j)$, and  (\ref{EHpm2}) takes the usual
form
\begin{equation}\label{EHpm21/2}
H^{-}_{2\bn}=U+\sum_{k<l}
J(\overline{\overline{\tau_{kl}}})\bs_k\bdot\bs_l.
\end{equation}
Although the same form is frequently used \cite{MANOUSAKIS} with
the spin operators for $s\neq 1/2$, these formal generalizations
do not preserve the original physical meaning of the
Heisenberg-Dirac hamiltonian: the resulting operator cannot be
derived from the pure orbital interaction of the identical
particles (for higher spin the transpositions cannot be expressed
by the spin matrices in the same form). Even for $s=1/2$, the
interaction coefficients can be independently chosen for any pair
of sites only for the occupation numbers $n_i\leq 1$; in other
cases, they must be same over the same double coset of $S^\bn_N$,
while the relevant space is only subspace of the total spin space
$\cH^N_s$, which can be easily found with help of the subgroup
projector $S^\bn_N(d\otimes D^{(-)})$. Indeed, using the direct
product factorization of the group $S^\bn_N$, the projector can be
written in the form: $S^\bn_N(d\otimes
D^{(-)})=\otimes^{\mod{o}}_{i=1}S_{n_i}(d\otimes D^{(-)})$. Each
of the factors may be straightforwardly found; moreover, only the
generating transpositions may be involved \cite{YI-GP}. This
simple restriction to the relevant space may be used to reduce the
time in various numerical calculations.

Finally, let it be emphasized that the general form of the
hamiltonian acting in the space of the ground representation of
the symmetry group (thus commuting with it), given by the theorem
\ref{Tinv} is the result important independently of the
Dirac-Heisenberg hamiltonian. Indeed, this situation occurs in the
context of the single particle approximations \cite{STERN}, e.g.
when the tight-binding electronic levels, spin waves or normal
vibrations modes are calculated: then the symmetry group action
can be factorized onto the permutational part $D^{{\rm P}}(\bG)$
and interior part $D^{{\rm int}}(\bG)$. The later is related to
the phenomena considered (this is polar and axial vector
representation of the group in the case of normal modes and spin
waves, and the representation carried by the atomic orbitals from
the same site in the electronic tight binding calculations). The
former describes the geometry of the system showing how the
transformations of the group map one site into another, and this
is always the ground representation induced from the site
stabilizer. Again the theorem \ref{Tinv} can be used to find the
general form of the hamiltonian, restricting possible theoretical
models and enabling further exact simplifications along to these
presented in the context of the Dirac-Heisenberg problem.


\end{document}